# The Fe-Ni phase diagram and the Earth's inner core structure

Liangrui Wei[1], Kai-Ming Ho[2], Renata M. Wentzcovitch[3-6], and Yang Sun[1]

[1]*Department of Physics, Xiamen University, Xiamen 361005, China*
[2]*Department of Physics, Iowa State University, Ames, IA 50011, USA*
[3]*Department of Applied Physics and Applied Mathematics, Columbia University, New York, NY 10027, USA*
[4]*Department of Earth and Environmental Sciences, Columbia University, New York, NY, 10027, USA*
[5]*Lamont–Doherty Earth Observatory, Columbia University, Palisades, NY, 10964, USA*
[6] *Data Science Institute, Columbia University, New York, NY 10027, USA*
(Dated: March 27, 2025)

The Fe-Ni alloy is believed to be the main component of Earth's core. Yet, a comprehensive understanding of phase equilibria near the melting point of this alloy under core conditions is still lacking, leaving the effect of nickel inconclusive. Using *ab initio* simulations, we computed Gibbs free energy and phase diagram for liquid and solid solutions of the Fe-Ni alloy under conditions close to the inner core, considering inner-shell electron contributions and non-ideal mixing effects. The Fe-Ni phase diagram provides crucial insights for understanding previous experimental observations and crystallization simulations of the Fe-Ni alloy under core conditions. It also presents new scenarios for inner core structures, suggesting *bcc*-liquid coexistence at the inner core boundary and the possibility of multi-layer structures consisting of *bcc-hcp* composites within the inner core. Our work clarifies nickel's substantial impact on the inner core structure, providing new constraints for the study of core's composition and formation.

The Earth's core consists of a liquid outer core (OC) and a solid inner core (IC), primarily made up of iron-nickel alloys and light elements [1,2]. Understanding the crystalline phases in the solid inner core is essential for determining the partitioning of light elements and the core's seismic properties. Traditionally, based on the phase diagram of pure iron, the inner core is thought to predominantly have a hexagonal close-packed (*hcp*) structure [3–5]. However, because the free energy of the body-centered cubic (*bcc*) phase can be comparable to that of the *hcp* phase under core conditions, the possibility of a *bcc* inner core has also been proposed [6,7].

Elements alloyed with iron can influence the stability of crystalline phases in the IC. While the exact concentration of light elements in the core remains debated [8–13], nickel is present in the OC and IC, with estimated concentrations ranging from 5% to 15% based on cosmochemical and geochemical models [1,14]. Low-pressure, low-temperature experiments have shown that nickel can enhance the stability of the face-centered cubic (*fcc*) phase over the hexagonal close-packed (*hcp*) phase in iron [15–17]. However, at high temperatures similar to core conditions, nickel was theoretically found to stabilize the *hcp* phase over the *fcc* phase in iron [18]. Given that *fcc* iron is metastable relative to *hcp* iron, the *fcc* phase of the Fe-Ni alloy has not been seriously considered for the IC structure [18]. The competition between *fcc*, *hcp*, and *bcc* phases in the Fe-Ni alloy under IC conditions remains largely unresolved.

The *bcc* Fe-Ni alloy was first observed experimentally at pressures above 225 GPa and temperatures exceeding 3400 K [19]. However, subsequent experiments did not confirm the existence of the *bcc* phase in Fe90Ni10 at 2,730 K and 250 GPa [20]. More recently, an ordered *bcc* phase (B2) was found to coexist with the *hcp* phase in Fe93Ni7 alloy at 2,970 K and 186 GPa [21]. Additionally, recent *ab initio* molecular dynamics (AIMD) simulations have revealed a *hcp-bcc* mixed structure that crystallized in Fe85Ni15 melts under IC conditions [22]. However, because the simulation involved large supercooling and rapid crystallization, it remains unclear whether the *hcp-bcc* mixed structure is thermodynamically stable. To fully understand the crystalline structure of the IC, it is crucial to compute the Fe-Ni phase diagram, accounting for the competition among the *bcc*, *hcp*, and *fcc* phases and their dependence on nickel concentrations under IC conditions.

Despite its significance, theoretically exploring the stability of the Fe-Ni alloy at IC conditions is challenging. Previous *ab initio* calculations of Fe-Ni alloys were conducted at core pressures but at 0 K, neglecting high-temperature effects [23]. Anharmonic vibrational effects at high temperatures play a critical role in the free energy and phase stability, especially for the *bcc* phase of iron at IC conditions. Additionally, the Fe-Ni alloy is a mixed system, meaning that ordered configurations alone cannot fully capture its free energy [7,24,25]. For example, the stability of *fcc* and *hcp* phases in the inner core is sensitive to the chemical disorder between Fe and Ni atoms [18]. Static calculations have already demonstrated that certain randomly mixed Fe-Ni configurations can stabilize the dynamic behavior of the *bcc* phase at IC pressures [26].

In this work, we use a hybrid approach that combines AIMD and Monte Carlo (MC) simulations to investigate the mixing effects of Fe and Ni in the solid phases. By calculating the *ab initio* mixing free energy, we construct the Fe-Ni solid-liquid phase diagram under inner core conditions. This phase diagram defines the stability of various solid and liquid solutions, enabling us to identify the stable solid phases in the IC based on temperature and nickel composition.

To obtain the Fe-Ni phase diagram, it is essential to compute the mixing enthalpy for the non-ideal solution phases under high-pressure and high-temperature conditions of IC. The multiple atomic configurations of solid solutions were sampled by using hybrid MC+AIMD simulations. Figure 1(a) compares MC+AIMD and pure AIMD for Fe$_{90}$Ni$_{10}$. Both simulations were initialized with the same randomly distributed FeNi configuration in a bcc lattice. Due to the swap mechanism, the atomic configurations in MC+AIMD effectively change, approaching the low-energy states. This results in a lower averaged potential energy in the MC+AIMD simulation by 17 meV/atom compared to the one from AIMD. Statistical analysis in Supplementary Text suggests that the error in enthalpy obtained from the MC+AIMD simulations is less than 2 meV/atom. Figure 1(b) shows dramatic structural differences between MC+AIMD and pure AIMD simulations based on the Ni-Ni pair correlation functions. Compared to the pure AIMD results, the MC+AIMD simulation leads to an enhanced first peak at 2.1Å and a third peak at 5.3 Å in g$_{Ni-Ni}$(r). The second peak of g$_{Ni-Ni}$(r) is shifted towards longer distances in the MC+AIMD result. These differences suggest that the Ni-Ni pairs in the MC+AIMD simulation exhibit short- to medium-range orders rather than a random distribution. To understand the spatial orders, we connect the Ni-Ni bonds using a cutoff value of first minimal in g$_{Ni-Ni}$(r) and analyze the size distribution of Ni clusters in Fig. 1(c). Compared to the randomly distributed Ni atoms in the same lattice, the configurations sampled by MC+AIMD show a lower probability of forming isolated Ni atoms but a higher probability of forming large clusters. We plot one snapshot in Fig. 1(d), which contains a large cluster of 14 Ni atoms. The Ni atoms form an interconnected framework, leading to a non-uniform Ni distribution in the lattice, as depicted by the coarse-grained distribution in Fig. 1(d). Thus, the MC+AIMD simulations demonstrate that a small amount of Ni tends to form non-ideal and non-uniform solid solutions in the Fe's lattice under IC conditions.

Based on the hybrid MC+AIMD simulations, the mixing enthalpy can be computed for all three solid phases. To obtain the mixing enthalpy at a consistent pressure and temperature of 323 GPa and 6000 K, we perform the simulations using the NVT ensemble with three different volumes near the target pressure for each phase, shown in Supplementary Materials Fig. S2. We mainly focus on the Fe-rich compositions relevant to the Earth's core. Figure 2(a) illustrates the mixing enthalpy as a function of Ni concentration for the *hcp*, *fcc*, *bcc*, and liquid solutions from hybrid MC+AIMD simulations at 323 GPa and 6000 K. The mixing enthalpy of the liquid phase is higher than that of all three solid phases. The *bcc* solution exhibits the lowest mixing enthalpy among the solid phases, while the *hcp* and *fcc* phases show similar mixing enthalpies. The mixing enthalpy of solid solutions at low Ni compositions (<10%) deviates from the usual regular solution model. In particular, the bcc solution shows a negative mixing enthalpy at 5 at.%, suggesting that mixing is energetically more favored than pure elemental phases at low Ni compositions, even if the entropy is not considered. The Redlich-Kister polynomial functions can fit these data well with errors less than 2 meV/atom. Martorell et al. previously examined the enthalpy of Fe-Ni alloys with static calculation at 0 K and 360 GPa [23]. By replotting the enthalpy data from Ref. [23] using the definition in Eqn. (3), similar trends can be observed between Figs. 2(a) and 2(b), consistently showing that the bcc solution has the lowest mixing enthalpy and negative values at low Ni compositions.

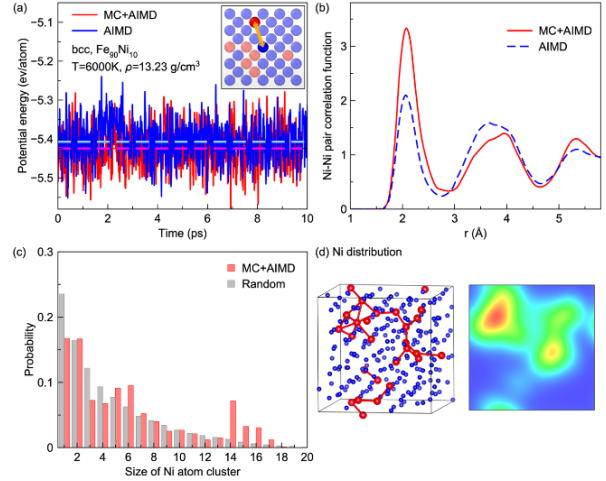

FIG. 1. MC+AIMD simulation of Fe$_{90}$Ni$_{10}$ solid solutions. (a) Potential energy as a function of time for Fe$_{90}$Ni$_{10}$ bcc phase from the MC+AIMD (red) and pure AIMD simulations (blue), starting from the same initial configuration. The simulation was carried out with fixed temperature and volume. The dashed lines are averaged potential energies for two simulations. The inset shows the schematic of the MC swap where two atoms of different types (shown in different colors) are randomly selected and swapped to sample the configurational space, with one swap attempt highlighted by the arrow. (b) The Ni-Ni pair correlation function from two simulations in (a). (c) The probabilities of Ni atoms forming different size of clusters. The statistical data of random distribution was obtained by distributing atoms randomly in the perfect lattice 10,000 times. The bond length threshold of forming a cluster is 2.9 Å, which is the first minimal of the Ni-Ni pair correlation function. (d) The left panel shows an atomic configuration with a Ni cluster of 14 atoms (longest chain) from MC+AIMD simulation. Red is Ni, while blue is Fe. The size of Fe atoms is reduced for clarity. Right panel shows the coarse-grained Ni distribution by a Gaussian smearing scheme as $D(\vec{r}) = \sum_i \left(\frac{1}{2\sigma^2\pi}\right)^{3/2} e^{-(\vec{r}-\vec{r}_i)^2/2\sigma^2}$, where $\vec{r}_i$ is the position of Ni atoms. $\sigma$ is set to 1.23 Å so that the full width at half maximum of the Gaussian function equals the first minimal of Ni-Ni pair correlation function.

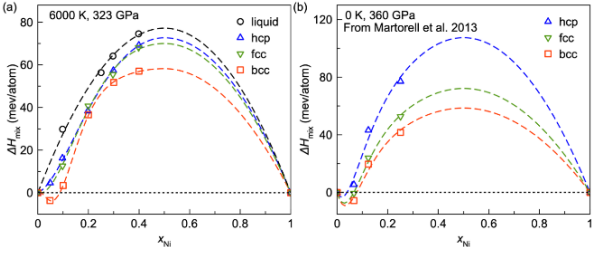

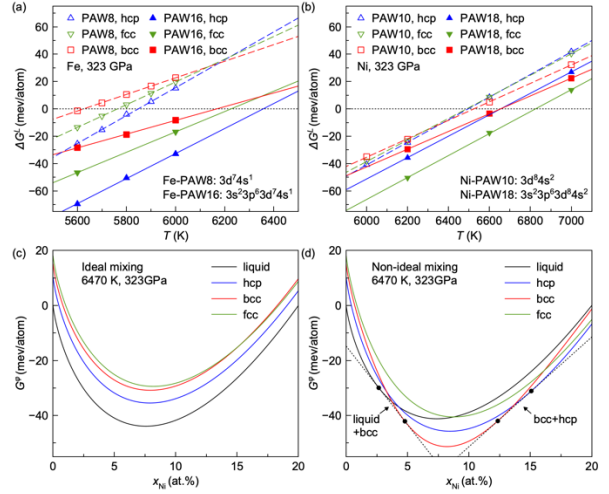

FIG. 2. **Mixing Enthalpy for Fe$_{1-x}$Ni$_x$ phases.** (a) The mixing enthalpy from our simulations at 323 GPa and 6000 K. The liquid data is from direct AIMD simulations. The data of solid solutions were obtained via hybrid MC+AIMD simulations. The dashed lines are the fitting results with the Redlich-Kister polynomial functions. The fitting parameters are included in Supplementary Text. (b) Mixing enthalpies at 0 K and 360 GPa from Ref. [23].

FIG. 3. *Ab initio* Gibbs free energy of pure and solutions phases. (a) The free energy of pure Fe phases, referenced to the liquid data at 323 GPa near the melting temperatures for Fe. The bcc and hcp data are from [24]. (b) The free energy of pure Ni, referenced to the liquid data at 323 GPa near the melting temperatures for Ni. The data obtained via PAW10 are from [22]. (c) Free energy of Fe-Ni phases from the ideal solution model, referenced to the liquid phase at 0% and 20% Ni. (d) Free energy of Fe-Ni phases from the regular solution model. The dashed lines represent the common tangent lines that define the solidus and liquidus compositions.

In addition to the mixing enthalpy, the Gibbs free energy of the pure end members is also crucial for the phase diagram calculations. We employ the recently developed free energy methods and previous *ab initio* datasets [22,24] to complete the free energy calculations of the liquid, *hcp*, *fcc*, and *bcc* phases for both Fe and Ni near their melting points at 323 GPa. We used the two sets of PAW potentials to calculate the free energy of the Fe and Ni end members, which was found to be important recently [24,25,42]. The relative energies w.r.t the liquid phases are shown in Figs. 3(a) and (b) for Fe and Ni, respectively. In the case of Fe in Fig. 3(a), *hcp* is the most stable phase, while *fcc* and *bcc* are metastable. Including $3s^23p^6$ inner-shell electrons for Fe increases melting point by approximately 500 K but does not change the relative phase stability among *hcp*, *fcc*, and *bcc* phases. In the case of Ni shown in Fig. 3(b), the $3s^23p^6$ electrons increase the melting point and modify the relative phase stability. Without $3s^23p^6$ electrons, *hcp*, *fcc*, and *bcc* phases have very similar free energies, with *bcc* as the most stable phase. However, when the $3s^23p^6$ electrons are considered, *fcc* becomes the most stable phase. The inner-shell electrons affect the relative phase stabilities for Fe and Ni. Thus, even though including $3s^23p^6$ inner-shell electrons increases the computational cost by a factor of 10, they are still necessary for a comprehensive description of phase competition for pure Fe and Ni systems under the large pressures of the IC. Figs. 3(a) and (b) suggest that the significantly higher melting temperature of Ni compared to Fe persists with different PAW potentials, as shown earlier [22]. Combining the pure free energy data and the mixing enthalpy, we can compute the Gibbs free energy of mixing in the regular solution model based on Eqn. (4). The Gibbs free energy of mixing was also compared with and without contributions of $3s^23p^6$ electrons. As shown in Supplementary Materials Fig. S3, including $3s^23p^6$ electrons systematically decreases the Gibbs free energy of mixing for all four phases, further stabilizing the mixture w.r.t the decomposition. Using the liquid phase as the reference, the relative Gibbs free energy of mixing changes differently for different phases with two sets of the PAW potentials, as shown in Supplementary Materials Fig. S4. Thus, the inclusion of $3s^23p^6$ electrons is necessary to correctly describe the competition among Fe-Ni solutions.

The comprehensive Gibbs free energy data of Fe-Ni alloys allow us to study the phase competition under IC

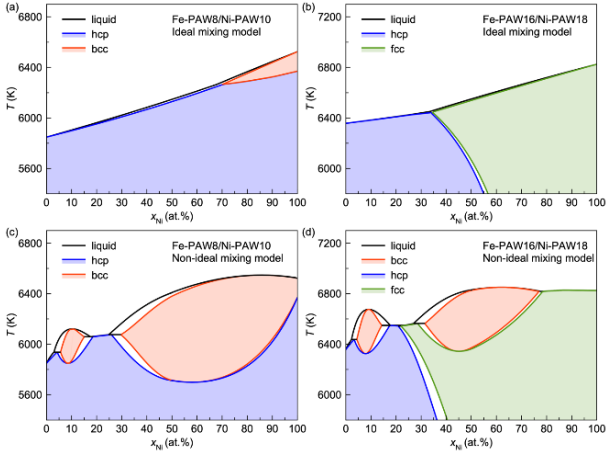

FIG. 4. Fe$_{1-x}$Ni$_x$ phase diagrams at 323 GPa using different models. (a) Ideal solution model without $3s^2 3p^6$ electron contributions. (b) Ideal mixing model with $3s^2 3p^6$ electron contributions. (c) Regular solution model without $3s^2 3p^6$ electron contributions. (d) Regular solution model with $3s^2 3p^6$ electron contributions.

conditions. In Fig. 3(c) and (d), we compare the Gibbs free energy among four Fe-Ni phases at 6470 K and analyze the differences between the ideal and the regular solid solution models. In the ideal mixing model shown in Fig. 3(c), the liquid phase is the lowest-energy state for 0% < $x_{Ni}$ < 20%. The small Gibbs free energy differences w.r.t the other phases of ~ 20 meV/atom, are similar or even smaller than the mixing enthalpy differences among the four phases, making the ideal mixing model unsuitable. As shown in Fig. 3(d), the regular solution model dramatically changes the relative Gibbs free energies. The common tangent lines can provide phase boundaries and the equilibrium phases at different compositions. It suggests that at 6470 K and 323 GPa, the liquid phase is stable for $x_{Ni}$ < 2.6%, and the liquid and $bcc$ phases coexist in equilibrium for 2.6% < $x_{Ni}$ < 4.8%. For 4.8% < $x_{Ni}$ < 12.3%, the $bcc$ phase is stable; for 12.3% < $x_{Ni}$ < 15%, the $bcc$ coexists with the $hcp$, and for $x_{Ni}$ > 15%, the $hcp$ is the only stable phase. Therefore, the non-ideality of the solid solution plays a crucial role in the phase stability of the Fe-Ni alloy at IC conditions. Besides, Fig. 3(d) indicates that the $bcc$ and $hcp$-$bcc$ mixtures can be stabilized at different temperatures, pressures, and compositions. We obtain the phase diagram by optimizing the free energy data w.r.t the phase compositions at various temperatures.

Figure 4 presents four phase diagrams illustrating the phase relationships among $bcc$, $hcp$, $fcc$, and liquid phases for the Fe$_{1-x}$Ni$_x$ alloys at 323 GPa, constructed using different solution models and PAW potentials. Across all phase diagrams, $hcp$ remains the stable phase at low nickel concentrations ($x_{Ni}$ < 3%), and adding nickel raises the melting temperature. This phenomenon can be attributed to two factors: (1) Ni has a higher melting temperature than Fe, and (2) the mixing enthalpy of the liquid phase is higher than that of all solid phases, further increasing the free energy of the liquid solution relative to the solid phase. There are notable differences among these four diagrams. Comparing Fig. 4(a) and (b) indicates that the inner-shell electrons $3s^2 3p^6$ largely stabilize the $fcc$ phase in Ni-rich compositions. The difference between Fig. 4(a) and (c) indicates that the regular solution model stabilizes the $bcc$ phase near the liquidus lines. The phase diagram in Fig. 4(d) includes the effects of inner-shell electrons and non-ideal mixing, thus providing the highest accuracy Fe-Ni phase diagram at IC conditions. It shows that the $bcc$ phase is stable within ~400 K below the liquidus lines in the range of 3%-18% and 28%-78% Ni compositions. At lower temperatures, the $hcp$ phase is favored in the Fe-rich composition, while the $fcc$ phase is stable for Ni-rich compositions. The stability field of the $bcc$ phase in Fe-rich compositions remains robust across different PAW potentials, as shown by the comparison between Fig. 4(c) and 4(d).

These phase diagrams provide helpful information to understand previous experiments and simulations. For instance, there were failed attempts to observe the $bcc$ phase in Fe$_{90}$Ni$_{10}$ alloy at 340 GPa and 4700 K [43]. Based on the phase diagram in Fig. 4(d), temperatures above 6300 K are required at 323 GPa to observe the $bcc$ phase. Thus, 4700 K is still too low to observe the $bcc$ near 340 GPa. The $bcc$ phase crystallized simultaneously in previous AIMD simulation with PAW10 potential for Ni at 5000 K and 323 GPa [22]. However, in Fig. 4(c), only the $hcp$ phase is stable in this region. The crystallization observed in [22] is attributed to the notably high nucleation rate of the metastable $bcc$ phase, suggesting that Ni likely follows a two-step nucleation mechanism, i.e., the metastable phase forms initially in the nucleation process, as observed for Fe under inner core conditions [44]. The phase diagram in Fig. 4(c) also helps explain the mixed $hcp$-$bcc$ phase crystallized in Fe$_{85}$Ni$_{15}$ alloy [22]. With 15% Ni, the phase diagram shows a stable region of $hcp$-$bcc$ coexistence at ~6000 K. Under 310 GPa and 5000 K conditions used in [22], while the $hcp$ should be the stable phase for the Fe$_{85}$Ni$_{15}$ alloy, the $bcc$ phase will likely have a higher nucleation rate to start the crystallization. Therefore, the system starts the nucleation process with the $bcc$ phase and grows an $hcp$-$bcc$ mixture to reduce the free energy. Such a mixture should also be a metastable phase, promoting the crystallization process, and could be the intermediate phase of two-step nucleation for the Fe$_{85}$Ni$_{15}$ alloy under IC conditions.

There are still limitations in present phase diagram calculations. The length scale in our current MC + AIMD simulations is restricted by the computational resources required for $ab$ $initio$ calculations. While the statistical error in enthalpy calculations was analyzed as insignificant, it remains unclear whether more significant medium-range ordering could emerge in larger-scale simulations [45,46]. Additionally, the mixing entropy in the regular solution model is the same as in the ideal solution model, disregarding additional vibrational entropy changes arising from alloying [47,48]. These limitations highlight the urgent need for a Fe-Ni interatomic potential with *ab initio* accuracy suitable for IC conditions to extend the simulation scales. The recent development of deep-learning techniques could provide a feasible approach to addressing such issues in future studies [42,49].

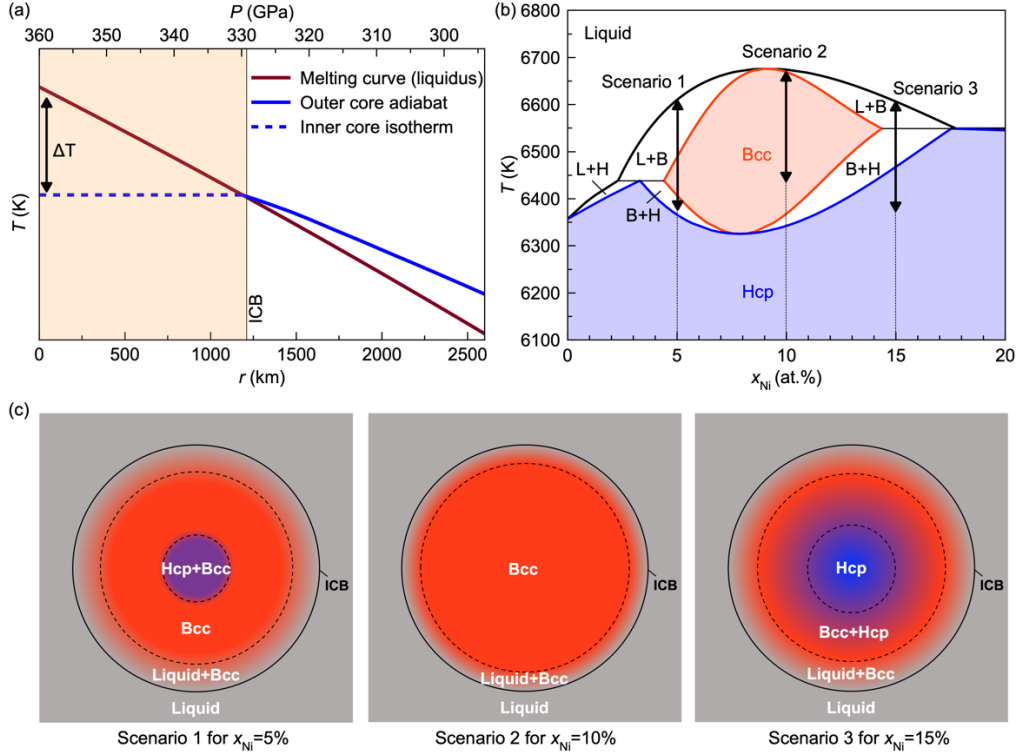

FIG. 5. Inner core structure implied by the Fe-Ni phase diagram. (a) Schematic of core's thermal structure. Outer core adiabats are anchored to the melting curve of the core's material at the ICB. The inner core is approximated as isothermal [50,51]. The arrow indicates the temperature difference between the *core* material's melting point and the inner core's temperature. (b) The Fe-Ni phase diagram for Fe-rich compositions relevant to the inner core. The arrow indicates a temperature span of 250 K from the ICB to the ICC at three different Ni compositions. (c) Different inner core structures based on the stable phases at Ni compositions of 5%, 10%, and 15%, deduced from the phase diagram in (b).

Based on the Fe-Ni phase diagram (Fig. 4(d)) and disregarding the effects of light elements for the moment, we can ponder on the effect of Ni on the present inner core structure - assuming it is in thermochemical equilibrium everywhere. There is a complex competition between *hcp*, *bcc*, and their mixtures in Fig. 4(d) for 5% < $x_{Ni}$ < 15%, the composition range relevant to the core [1,14]. Based on the current understanding of the core's thermal structure [50], the temperature at the inner core boundary (ICB) is at the intersection of liquidus line of core materials and the liquid core adiabat.

The inner core should be isothermal due to its large thermal conductivity [50,51]. The pressure increase toward the inner core center (ICC) should raise the liquidus temperature of the alloy. Here, we assume this increase in liquidus temperature, $\Delta T$, is similar to the increase in melting temperature of Fe with pressure, i.e., approximately 200-300 K from 330 GPa to 360 GPa [24,52]. As indicated in Fig. 5(b) and 5(c), the multi-phase equilibrium state changes with varying $\Delta T$, i.e., depth and initial alloy composition, forming a multi-layer solid core. For estimated compositions 5% < $x_{Ni}$ < 15% [1,14], we show the inferred inner core structures in Fig. 5(c). If $x_{Ni}$ = 5%, one should expect a large range of liquid+*bcc* mixture near the ICB. With increasing depth, a layer of pure *bcc* phase stabilizes, and a mixture of *hcp*+*bcc* stabilizes at the center. Small changes in $x_{Ni}$ can change the thickness of these layers. If $x_{Ni}$ =10%, the liquid+*bcc* layer near the ICB becomes relatively thinner, with only a single *bcc* phase in the inner core. If $x_{Ni}$ =15%, the pure *bcc* phase domain should disappear, and a pure *hcp* phase should emerge at the ICC. The phase diagram in Fig. 5(b) suggests that a liquid+*bcc* mixture at the ICB should always exist for 5% < $x_{Ni}$ < 15%, as shown in Fig. 5(c). However, the solid structure in the IC is sensitive to the Ni concentration. Within this alloy concentration range, the *bcc* phase always exists somewhere while *hcp* may not (as in Scenario 2). The *hcp* becomes the only solid phase for $x_{Ni} < 3\%$ or $x_{Ni} > 18\%$.

The inner core structure deduced from the Fe-Ni phase diagram is, therefore, much more complex than the typical structure derived from pure Fe. Since the phase diagram of the Fe-Ni alloy is quite different from that of pure Fe at relevant conditions, the presence of Ni in the core might affect the current understanding of light element partitioning between IC and OC [8]. The Fe-Ni alloy's distinct crystallization kinetics may provide new insights into the inner core formation and evolution [44,53,54].

In summary, using regular solution models, we computed the *ab initio* phase diagram of the Fe-Ni system under IC conditions. We employed a hybrid MC+AIMD simulation to explore different configurations of $Fe_{1-x}Ni_x$ solid solutions, a task that is not feasible with conventional AIMD. The simulation reveals that Ni atoms cluster and form non-ideal and non-uniform solid solutions with Fe at core conditions. By comparing phase diagrams computed with different accuracy levels, we demonstrate that the effect of inner-shell electrons and non-ideal mixing are critical for accurately determining the Fe-Ni phase diagram at core conditions. At 323 GPa and $4\% < x_{Ni} < 18\%$ or $28\% < x_{Ni} < 78\%$, *bcc* is the stable phase for temperatures down to $\sim 400$ K below the liquidus line. At lower temperatures, *hcp* is stable in Fe-rich compositions, while *fcc* is stable in Ni-rich compositions. The phase diagram provides useful insights for understanding previous experimental observations and crystallization simulation for the Fe-Ni alloy under core conditions. It also suggests new scenarios for IC structures far more complex than usually assumed with pure Fe phases. The ICB should exhibit liquid-*bcc* coexistence. The IC can have multiple layers composed of *hcp*, *bcc*, and their mixture, which highly depends on the Ni's composition. The present study emphasizes the crucial effect of Ni on the IC structure. It also paves the way for future calculations of thermodynamic properties of multicomponent systems with Fe, Ni, and light elements.

## Methods

*Ab initio* calculations were performed with the Vienna *ab initio* simulation package (VASP) [27]. The electron-ion interaction was described using the Projector Augmented-Wave (PAW) method [28], and the exchange-correlation energy was treated with the Generalized Gradient Approximation (GGA) in the Perdew-Burke-Ernzerhof (PBE) form [29]. The electronic entropy was described by the Mermin functional [30,31]. The electronic temperature in the Mermin functional is kept the same as the ionic temperature. Two sets of PAW-PBE potentials were used for AIMD and MC simulations: PAW8 with $3d^74s^1$ valence electrons for Fe and PAW10 with $3d^84s^2$ for Ni. PAW16 with $3s^23p^63d^74s^1$ valence electrons for Fe and PAW18 with $3s^23p^63d^84s^2$ for Ni were also employed to improve the DFT accuracy via free energy perturbation (FEP) [24,25,32]. The plane-wave cutoff energies were set to 400 eV for PAW8/PAW10 and 750 eV for PAW16/PAW18. Supercells containing 288, 250, 256, and 256 atoms were used for the hcp, bcc, fcc, and liquid phases, respectively. The $\Gamma$ point was used to sample the Brillouin zone in AIMD and MC simulations. A dense Monkhorst-Pack k-point mesh of $2 \times 2 \times 2$ was adopted for all phases to achieve high DFT accuracy in the FEP calculations.

We employed the MC swap algorithm for the solid phases (bcc, hcp, fcc) to sample mixing configurations of Fe-Ni solid solutions. Fe and Ni atoms were randomly interchanged on the fly during the AIMD simulation. The acceptance probability of the trial swap is determined based on the Metropolis algorithm with the probability of $\min(1, \exp(-\Delta E/k_B T))$, where $\Delta E$ represents the total energy change due to the swap. One swap was attempted every 100 MD steps with approximately 80% acceptance rates on average. The Nosé-Hoover thermostat [33] was used to control the temperature, and a time step of 1.0 fs was employed to integrate Newton's equations of motion in the MC+AIMD simulations. This hybrid approach can significantly accelerate the equilibration process and enhance the configurational diversity of the simulated ensemble for solid solutions [34–36].

For completeness, we briefly review the thermodynamics of binary systems for phase diagram construction. When Fe and Ni are mixed to form a solution phase $\varphi$ at constant temperature and pressure, the change in Gibbs free energy during mixing, referred to as the Gibbs free energy of mixing, is expressed as

$$\Delta G_{mix}^\varphi = \Delta H_{mix}^\varphi - T\Delta S_{mix}^\varphi, \quad (1)$$

where $\Delta H_{mix}^\varphi$ is the mixing enthalpy and the $\Delta S_{mix}^\varphi$ is the mixing entropy. In an ideal solution model, $\Delta H_{mix}^\varphi = 0$ and the entropy of mixing is given by

$$\Delta S_{mix}^\varphi = -k_B T[x\ln x + (1-x)\ln(1-x)], (2)$$

where $k_B$ is the Boltzmann constant and $x$ is the composition of nickel. In a realistic solution, $\Delta H_{mix}^\varphi \neq 0$. So a regular solution model [37] with non-ideal mixing enthalpy is needed, represented as,

$$\Delta H_{mix}^\varphi = H_{Fe_{1-x}Ni_x}^\varphi - (1-x)H_{Fe}^\varphi - xH_{Ni}^\varphi, (3)$$

where $H_{Fe}^\varphi$ and $H_{Ni}^\varphi$ are the enthalpy of pure Fe and Ni in phase $\varphi$. The regular solution model often captures the detailed features of the phase diagram as long as $\Delta H_{mix}^\varphi$ is calculated accurately [38]. The Gibbs free energy for the solution phase is

$$G^\varphi(x,T,P) = \Delta G_{mix}^\varphi + (1-x)G_{Fe}^\varphi + xG_{Ni}^\varphi (4)$$

where $G_{\text{Fe}}^{\varphi}$ and $G_{\text{Ni}}^{\varphi}$ are the free energy of pure Fe and Ni, respectively. $\varphi$ denotes the liquid, *hcp*, *fcc* or *bcc* phase in this study.

The binary phase diagram of a multi-phase mixture is determined by minimizing the total Gibbs free energy of the assemblage. This process can be effectively addressed using the CALPHAD (CALculation of PHAse Diagrams) approach [39]. The Gibbs free energy data of different Fe-Ni phases computed from the first-principles simulations were fitted by the Redlich-Kister (RK) expression [40] as

$$G^{\varphi}(x, T, P) = \sum_{i=\text{Fe,Ni}} x_i G_i^{\varphi} + k_B T \sum_i x_i \ln x_i + x_{\text{Fe}} x_{\text{Ni}} \sum_{v=0}^{n} L_{\varphi}^{v} (x_{\text{Fe}} - x_{\text{Ni}})^v, \quad (5)$$

where $L_{\varphi}^{v}$ are the fitting parameter. The multi-phase free energy minimization was solved by computing the convex hull via the *Pycalphad* software [41].

## Acknowledgements

Work at Xiamen University was supported by the National Natural Science Foundation of China (Grants Nos. T2422016 and 42374108) and Fujian Provincial Natural Science Foundation of China (Grant No. 2024J09003). RMW acknowledges support from National Science Foundation (Grants Nos. EAR-2000850 and EAR-1918126). KMH acknowledges support from National Science Foundation (Grant No. EAR-1918134). S. Fang and T. Wu from the Information and Network Center of Xiamen University are acknowledged for their help with Graphics Processing Unit computing. The supercomputing time was partly supported by the Opening Project of the Joint Laboratory for Planetary Science and Supercomputing, Research Center for Planetary Science, and the National Supercomputing Center in Chengdu (Grants No. CSYYGS-QT-2024-15).

## References

[1] F. Birch, Elasticity and constitution of the Earth's interior, J. Geophys. Res. **57**, 227 (1952).

[2] K. Hirose, S. Labrosse, and J. Hernlund, Composition and State of the Core, Annu. Rev. Earth Planet. Sci. **41**, 657 (2013).

[3] O. L. Anderson, Properties of iron at the Earth's core conditions, Geophys. J. Int. **84**, 561 (1986).

[4] G. Steinle-Neumann, L. Stixrude, R. E. Cohen, and O. Gülseren, Elasticity of iron at the temperature of the Earth's inner core, Nature **413**, 57 (2001).

[5] Y. Zhang et al., Collective motion in hcp-Fe at Earth's inner core conditions, Proc. Natl. Acad. Sci. **120**, e2309952120 (2023).

[6] L. Vočadlo, I. G. Wood, M. J. Gillan, J. Brodholt, D. P. Dobson, G. D. Price, and D. Alfè, The stability of bcc-Fe at high pressures and temperatures with respect to tetragonal strain, Phys. Earth Planet. Inter. **170**, 52 (2008).

[7] A. B. Belonoshko, T. Lukinov, J. Fu, J. Zhao, S. Davis, and S. I. Simak, Stabilization of body-centred cubic iron under inner-core conditions, Nat. Geosci. **10**, 312 (2017).

[8] K. Hirose, B. Wood, and L. Vočadlo, Light elements in the Earth's core, Nat. Rev. Earth Environ. **2**, 645 (2021).

[9] J. Liu, Y. Sun, C. Lv, F. Zhang, S. Fu, V. B. Prakapenka, C. Wang, K. Ho, J. Lin, and R. M. Wentzcovitch, Iron-rich Fe–O compounds at Earth's core pressures, The Innovation **4**, 100354 (2023).

[10] Y. Tian, P. Zhang, W. Zhang, X. Feng, S. A. T. Redfern, and H. Liu, Iron alloys of volatile elements in the deep Earth's interior, Nat. Commun. **15**, 3320 (2024).

[11] T. Liu and Z. Jing, Hydrogen and silicon are the preferred light elements in Earth's core, Commun. Earth Environ. **5**, 1 (2024).

[12] F. Sakai, K. Hirose, and G. Morard, Partitioning of silicon and sulfur between solid and liquid iron under core pressures: Constraints on Earth's core composition, Earth Planet. Sci. Lett. **624**, 118449 (2023).

[13] L. Yuan and G. Steinle-Neumann, Hydrogen distribution between the Earth's inner and outer core, Earth Planet. Sci. Lett. **609**, 118084 (2023).

[14] V. F. Buchwald, Handbook of iron meteorites. Their history, distribution, composition and structure, Ariz. State Univ. (1975).

[15] J. Lin, D. L. Heinz, A. J. Campbell, J. M. Devine, W. L. Mao, and G. Shen, Iron-Nickel alloy in the Earth's core, Geophys. Res. Lett. **29**, 109 (2002).

[16] T. Komabayashi, Phase Relations of Earth's Core-Forming Materials, Crystals **11**, 581 (2021).

[17] T. Komabayashi, K. Hirose, and Y. Ohishi, In situ X-ray diffraction measurements of the fcc–hcp phase transition boundary of an Fe–Ni alloy in an internally heated diamond anvil cell, Phys. Chem. Miner. **39**, 329 (2012).

[18] M. Ekholm, A. S. Mikhaylushkin, S. I. Simak, B. Johansson, and I. A. Abrikosov, Configurational thermodynamics of Fe–Ni alloys at Earth's core conditions, Earth Planet. Sci. Lett. **308**, 90 (2011).

[19] L. Dubrovinsky et al., Body-Centered Cubic Iron-Nickel Alloy in Earth's Core, Science **316**, 1880 (2007).

[20] T. Sakai, E. Ohtani, N. Hirao, and Y. Ohishi, Stability field of the hcp-structure for Fe, Fe-Ni, and Fe-Ni-Si alloys up to 3 Mbar, Geophys. Res. Lett. **38**, 2011GL047178 (2011).

[21] D. Ikuta, E. Ohtani, and N. Hirao, Two-phase mixture of iron–nickel–silicon alloys in the

Earth's inner core, Commun. Earth Environ. **2**, 225 (2021).

[22] Y. Sun, M. I. Mendelev, F. Zhang, X. Liu, B. Da, C.-Z. Wang, R. M. Wentzcovitch, and K.-M. Ho, Unveiling the effect of Ni on the formation and structure of Earth's inner core, Proc. Natl. Acad. Sci. **121**, e2316477121 (2024).

[23] B. Martorell, J. Brodholt, I. G. Wood, and L. Vočadlo, The effect of nickel on the properties of iron at the conditions of Earth's inner core: Ab initio calculations of seismic wave velocities of Fe–Ni alloys, Earth Planet. Sci. Lett. **365**, 143 (2013).

[24] Y. Sun, M. I. Mendelev, F. Zhang, X. Liu, B. Da, C. Wang, R. M. Wentzcovitch, and K. Ho, *Ab Initio* Melting Temperatures of Bcc and Hcp Iron Under the Earth's Inner Core Condition, Geophys. Res. Lett. **50**, e2022GL102447 (2023).

[25] T. Sun, J. P. Brodholt, Y. Li, and L. Vočadlo, Melting properties from *ab initio* free energy calculations: Iron at the Earth's inner-core boundary, Phys. Rev. B **98**, 224301 (2018).

[26] S. Chatterjee, S. Ghosh, and T. Saha-Dasgupta, Ni Doping: A Viable Route to Make Body-Centered-Cubic Fe Stable at Earth's Inner Core, Minerals **11**, 258 (2021).

[27] G. Kresse and J. Furthmüller, Efficient iterative schemes for *ab initio* total-energy calculations using a plane-wave basis set, Phys. Rev. B **54**, 11169 (1996).

[28] P. E. Blöchl, Projector augmented-wave method, Phys. Rev. B **50**, 17953 (1994).

[29] J. P. Perdew, K. Burke, and M. Ernzerhof, Generalized Gradient Approximation Made Simple, Phys. Rev. Lett. **77**, 3865 (1996).

[30] N. D. Mermin, Thermal Properties of the Inhomogeneous Electron Gas, Phys. Rev. **137**, A1441 (1965).

[31] R. M. Wentzcovitch, J. L. Martins, and P. B. Allen, Energy versus free-energy conservation in first-principles molecular dynamics, Phys. Rev. B **45**, 11372 (1992).

[32] R. W. Zwanzig, High-temperature equation of state by a perturbation method. I. Nonpolar gases, J. Chem. Phys. **22**, 1420 (1954).

[33] S. Nosé, A unified formulation of the constant temperature molecular dynamics methods, J. Chem. Phys. **81**, 511 (1984).

[34] M. Widom, W. P. Huhn, S. Maiti, and W. Steurer, Hybrid Monte Carlo/Molecular Dynamics Simulation of a Refractory Metal High Entropy Alloy, Metall. Mater. Trans. A **45**, 196 (2014).

[35] Z. Zhang, G. Csányi, and D. Alfè, Partitioning of sulfur between solid and liquid iron under Earth's core conditions: Constraints from atomistic

simulations with machine learning potentials, Geochim. Cosmochim. Acta **291**, 5 (2020).

[36] Y. Huang, M. Widom, and M. C. Gao, *Ab initio* free energies of liquid metal alloys: Application to the phase diagrams of Li-Na and K-Na, Phys. Rev. Mater. **6**, 013802 (2022).

[37] D. R. Gaskell and D. E. Laughlin, *Introduction to the Thermodynamics of Materials* (CRC Press, 2017).

[38] L. Kaufman and H. Bernstein, *Computer Calculation of Phase Diagrams* (Academic Press, New York, 1970).

[39] P. J. Spencer, A brief history of CALPHAD, Calphad **32**, 1 (2008).

[40] O. Redlich and A. T. Kister, Algebraic Representation of Thermodynamic Properties and the Classification of Solutions, Ind. Eng. Chem. **40**, 345 (1948).

[41] R. Otis and Z.-K. Liu, pycalphad: CALPHAD-based Computational Thermodynamics in Python, J. Open Res. Softw. **5**, 1 (2017).

[42] F. Wu, S. Wu, C.-Z. Wang, K.-M. Ho, R. Wentzcovitch, and Sun, Melting temperature of iron under the Earth's inner core condition from deep machine learning, Geosci. Front. **15**, 101925 (2024).

[43] S. Tateno, K. Hirose, T. Komabayashi, H. Ozawa, and Y. Ohishi, The structure of Fe-Ni alloy in Earth's inner core, Geophys. Res. Lett. **39**, 2 (2012).

[44] Y. Sun, F. Zhang, M. I. Mendelev, R. M. Wentzcovitch, and K.-M. Ho, Two-step nucleation of the Earth's inner core, Proc. Natl. Acad. Sci. **119**, e2113059119 (2022).

[45] J. Wang, P. Jiang, F. Yuan, and X. Wu, Chemical medium-range order in a medium-entropy alloy, Nat. Commun. **13**, 1021 (2022).

[46] Z. An et al., Negative enthalpy alloys and local chemical ordering: a concept and route leading to synergy of strength and ductility, Natl. Sci. Rev. **11**, nwae026 (2024).

[47] A. van de Walle and G. Ceder, The effect of lattice vibrations on substitutional alloy thermodynamics, Rev. Mod. Phys. **74**, 11 (2002).

[48] R. M. Wentzcovitch, J. F. Justo, Z. Wu, C. R. S. da Silva, D. A. Yuen, and D. Kohlstedt, Anomalous compressibility of ferropericlase throughout the iron spin cross-over, Proc. Natl. Acad. Sci. **106**, 8447 (2009).

[49] F. Wu, Y. Sun, T. Wan, S. Wu, and R. M. Wentzcovitch, Deep-Learning-Based Prediction of the Tetragonal → Cubic Transition in Davemaoite, Geophys. Res. Lett. **51**, e2023GL108012 (2024).

[50] R. A. Fischer, *Melting of Fe Alloys and the Thermal Structure of the Core*, in *Geophysical Monograph Series* (2016), pp. 1–12.

[51] M. Pozzo, C. Davies, D. Gubbins, and D. Alfè, Thermal and electrical conductivity of solid iron and iron–silicon mixtures at Earth's core conditions, Earth Planet. Sci. Lett. **393**, 159 (2014).

[52] D. Alfè, G. D. Price, and M. J. Gillan, Iron under Earth's core conditions: Liquid-state thermodynamics and high-pressure melting curve from ab initio calculations, Phys. Rev. B **65**, 165118 (2002).

[53] L. Huguet, J. A. Van Orman, S. A. Hauck, and M. A. Willard, Earth's inner core nucleation paradox, Earth Planet. Sci. Lett. **487**, 9 (2018).

[54] A. J. Wilson, D. Alfè, A. M. Walker, and C. J. Davies, Can homogeneous nucleation resolve the inner core nucleation paradox?, Earth Planet. Sci. Lett. **614**, 118176 (2023).